
%

\magnification \magstep1

\hoffset 1.5truecm
\hsize 16truecm \vsize 23truecm
\baselineskip 20pt \parskip5pt
\raggedbottom

\font\cs=cmcsc10

\def\bsquare{\vbox{\hrule width.7em height.7em}}
\def\cL{{\cal L}}
\def\intpar{\hfil\break\indent}
\def\Keq#1#2{\Bigl\langle #1_{K_1},#2_{K_2}\Bigr\rangle
=\Bigl\langle #1_{K_1\#_b \tilde K_2},#2_{K_2}\Bigr\rangle}
\def\lk{\ell\!k}
\def\M{{\cal M}}
\def\rtw{I_r(M)}
\def\section#1{\bigskip\penalty-100\centerline{\cs #1}\par}
\def\si{{\cal I}_r(\M)}


\rightline{hep-th/9302092}                       

\vfill

\centerline{\bf SURGICAL INVARIANTS OF FOUR-MANIFOLDS}

\smallskip

\centerline{BOGUS\L AW BRODA}

\bigskip

\noindent
{\cs Abstract.}
A new topological invariant of closed connected orientable
four-dimensional manifolds is proposed. The invariant,
constructed via surgery on a special link, is a
four-dimensional counterpart of the celebrated SU(2)
three-manifold invariant of Reshetik\-hin, Turaev and
Witten.

\vfill
\centerline{\cs February 1993}

\vfill

\centerline{REVISED}

\centerline{\cs March 1993}

\vfill
\footnote{}%
{1991 {\it Mathematics Subject Classification}.  Primary
57N13; Secondary 57M25,
\hfil\break
57N10, 57R65.
\intpar
\intpar
This research was supported by the Alexander von Humboldt
Foundation and the KBN grant no 202189101.
}

\vfill\eject

\section{Introduction}
The issue of topological classification of low-dimensional
manifolds, especially of the dimension {\it three} and {\it
four}, is one of the most challenging problems in modern
mathematics. One of the most spectacular events in topology
of three-dimensional manifolds took place in 1989, when a
new (numerical) topological invariant $\rtw$  of a closed
orientable three-dimensional manifold $M$, parametrized by
the integer $r$ ($r\geq2$), defined via {\it surgery} on a
framed link, was discovered. The idea is due to a physicist
[1], Edward Witten, an explicit construction to
mathematicians, Reshetikhin and Turaev [2], further
studies, to Kirby and Melvin [3], whereas a significant
simplification of the method to Lickorish [4]. The
invariant $\rtw$, known as the {\it
Reshetikhin-Turaev-Witten} (RTW) invariant, is also
frequently referred to as the {\it SU(2)-invariant} because
the {\it Kauffman bracket\/} it bases upon (denoted in
mathematical literature as `$\left<\;\right>$') formally
corresponds, in Witten's approach, to the average with
respect to the connection $\cal A$ (defined on a trivial
SU(2) bundle on the three-manifold $M$) modulo gauge
transformations, weighted by $\exp\left[ik{\rm
cs}(A)\right]$. Here $k$ is a positive integer, called the
{\it level}, and ${\rm cs}({\cal A})$ is the Chern-Simons
{\it secondary characteristic class}.  Incidentally, the
average is also denoted as `$\left<\;\right>$'. The
construction of the RTW invariant makes use of the {\it
fundamental theorem of surgery} of Lickorish and Wallace on
presentation of every closed connected orientable
three-manifold $M$ via surgery on a framed link in $S^3$,
and the {\it linear skein theory} associated with the
Kauffman bracket. The derivation of $\rtw$ is
combinatorial, and amounts to showing the invariance of
$\rtw$ with respect to the {\it Kirby moves}.

In four dimensions, there is a celebrated theorem of
Freedman [5] on classification of closed orientable {\it
simply-connected\/} four-manifolds, provided by the {\it
intersection form} $Q(\M)$ (and the Kirby-Siebenman
invariant $\alpha(\M)$). The intersection form $Q(\M)$
corresponds to, and for four-dimensional manifolds with
boundary defined via surgery on a link in $S^3$ is equal
to, the {\it linking matrix} $\lk$. The elements of the
symmetric matrix $\lk$, the {\it linking numbers} (with
framings on the diagonal), are the simplest numerical
invariants of a link. Therefore, one can ask the following
questions. Can one use some stronger (`non-abelian')
invariants of links, for example the Kauffman bracket
polynomial, to obtain some new `non-abelian' invariants of
four-dimensional manifolds?  Can one extend the idea of RTW
to the four-dimensional case?  Can one treat
simply-connected and non-simply-connected manifolds
uniquely? The answer to these questions seems to be
affirmative. Namely, we would like to propose a new
invariant of closed connected orientable four-manifolds,
defined via surgery on a {\it special\/} link in $S^3$.
Thus, we have succeeded in finding a quantity invariant
with respect to the four-dimensional version of the `Kirby
moves'. The idea as well as the construction resembles the
original one, proposed by RTW in the three-dimensional
case, whereas the four-dimensional version of the {\it
Kirby calculus} we need has been developed by
C\'esar~de~S\'a in [6].

\section{Construction of the invariant}
An arbitrary closed connected orientable
four-manifold $\M$ can be obtained via surgery in $S^3$
on a {\it special framed link} $\left(\cL,f\right)$ [6]. By
definition, the special framed link $\cL$ is a sum of
two sorts of knots
$$
\cL
=\bigcup_{i=1}^n K_i
\cup
\bigsqcup_{i=1}^{\dot n} \dot K_i,
$$
where $\bigl\{K_i\bigr\}_{i=1}^n$ are ordinary knots, and
$\bigl\{\dot K_i\bigr\}_{i=1}^{\dot n}$ are special ones.
The special knots are trivial (with zero framing), and
mutually unlinked unknots, and the whole link, when
regarded as a description of a three-manifold, represents a
connected sum of copies of $S^1\times S^2$.

Following the terminology of Lickorish [4], for a given
(fixed) integer $r$, $r\geq2$, let $\omega$ belonging to
the linear skein of the annulus ${\cal S}(S^1\times I)$ be
defined by
$$
\omega=\sum_{m=0}^{2r-2} \Delta_m S_m(\alpha),
$$
where the coefficients
$$
\Delta_m=
{(-1)^m \left(A^{2(m+1)}-A^{-2(m+1)}\right)
\over
A^2-A^{-2}},
$$
$S_m(\alpha)$ is the $m^{\rm th}$ {\it Chebyshev
polynomial\/} in the generator $\alpha$ of ${\cal
S}(S^1\times I)$, and $A$ is a primitive $8r^{\rm th}$
root of unity.

Introducing a ${\bf Z}_2$-gradation, we can decompose
$\omega$ as follows (compare [7])
$$
\omega=\omega^+ + \omega^-,
$$
where
$$
\omega^+=\sum_{m=0}^{r-1} \Delta_{2m} S_{2m}(\alpha),
$$
$$
\omega^-=\sum_{m=0}^{r-2} \Delta_{2m+1} S_{2m+1}(\alpha).
$$
The gradation used is provided by the power of $\alpha$, or
equivalently by the degree of the Chebyshev polynomial
$S_m(\alpha)$, or alternatively by the `spin' labeling
irreducible representations of the (quantum) SU(2) group.

\noindent
{\it Notation.}
We denote as $\tilde K$ the result of pushing a knot $K$
off itself (missing the rest of the link $L$) using the
framing $f$ of $K$, whereas as $K_1\#_b K_2$ a (band)
connected sum of the two knots $K_1$, $K_2$, where $b$ is
any band missing the rest of $L$. $\omega_K$ denotes
$\omega$ immersed in the plane as a regular neighbourhood
of $K$.

\noindent
{\it Proposition.}
Let $a^+$, $a^-$ be arbitrary complex numbers. Then we have
the following `Kirby calculus'
$$
\Keq{\alpha}{\omega},
$$
$$
\Keq{\alpha^2}{\left(a^+ \omega^+ + a^-\omega^-\right)}.
$$
The first equality expresses a standard property of
$\omega$ [4], whereas the second one follows from the
observation that the even element $\alpha^2$ respects the
gradation in all cablings (the Fenn-Rourke version of this
phenomenon is implicit in [7]).

\noindent
{\it Corollary.}
{}From the Proposition we can derive the following
`Kirby equalities'
$$
\Keq{\omega}{\omega},
$$
$$
\Keq{\omega^+}{\omega^+},
$$
$$
\Keq{\omega^+}{\omega}.
$$
Fortunately, the fourth lacking equality,
$$
\Keq{\omega}{\omega^+},
$$
is, in general, not true.

Henceforth, $U$ denotes a trivial (with zero framing)
unknot, $H_1$ and $\dot H_2$ are two components of the {\it
special Hopf\/} link $\cal H$, ordinary and special
respectively, ${\cal H}=H_1\cup\dot H_2$.

\noindent
{\bf Theorem.}
{\it
Let $n$ and $\nu$ be the dimension and nullity of the
linking matrix $\lk$.  Then
$$
\si=
{\left<\prod_{i=1}^n \omega_{K_i}^+
\prod_{i=1}^{\dot n} \omega_{\dot K_i}\right>
\over
\left< \omega_U^+ \right>^\nu
\left< \omega_{H_1}^+ \omega_{\dot H_2} \right>^{(n-\nu)/2}}
$$
is an invariant of (closed, connected, orientable four-manifold)
$\M=\M_\cL$, a complex number parametrized by the
integer $r$, $r\geq2$, independent of the choice of the
representative $\left(\cL,f\right)$.
}

Below, we give a list of all the allowable
`four-dimensional Kirby moves', so-called $\Gamma$-{\it
moves} [6]:

{\parskip0pt

\item{($a$)}
sliding one of the special knots over another special one;

\item{($b$)}
sliding one of the ordinary knots over one of the special ones;

\item{($c$)}
sliding one of the ordinary knots over another ordinary
one;

\item{($d$)}
introducing or deleting a special Hopf link;

\item{($e$)}
introducing or deleting a trivial unknot;

\item{($f$)}
isotoping the link picture in $S^3$.
}

\noindent
{\it Sketch of the proof.}
We should show that $\si$ is invariant with respect to all
the $\Gamma$-moves. $a$-, $b$- and $c$-invariance of $\si$
immediately follows from the Corollary. $d$-invariance is a
consequence of the following transformation rule of the
linking matrix $\lk$, accompanying the introduction of a
special Hopf link $\cal H$,
$$
\lk\longrightarrow\left(\matrix{\lk&0&0\cr
                                  0&0&1\cr
                                  0&1&0\cr}\right).
$$
Hence the corresponding shift of the dimension and nullity
of $\lk$
$$
\eqalign{n&\longrightarrow n+2\cr
       \nu&\longrightarrow \nu,}
$$
compensates the (factorized out) Kauffman bracket in the
numerator. Similarly, $e$-invariance corresponds to the
transformation rule
$$
\lk\longrightarrow\left(\matrix{\lk&0\cr
                                  0&0\cr}\right),
$$
and consequently the shift
$$
\eqalign{n&\longrightarrow n+1\cr
       \nu&\longrightarrow \nu+1,}
$$
also compensates the numerator. $f$-invariance directly
follows from fundamental properties of the Kauffman bracket and
the linking matrix $\lk$.
\hfil \bsquare

\noindent
{\it Remarks.}
(i) In the particular case of a {\it simply-connected}
$\M$, $\si$ simplifies to the exponent of a linear
combination of the signature and the Betti number;
(ii)~There is also a four-dimensional counterpart of the
invariant of Turaev and Viro, defined via triangulations by
Ooguri, Crane and Yetter [8].

\section{Acknowledgments}
The author would like to thank Professors
W.~B.~R.~Lickorish, J.~Przytycki and C.~Taubes for
inspiring discussions during his stay at the Newton
Institute, and especially J\'ozef Przytycki for further
correspondence. The author is very grateful to
Prof.~P.~Gilmer for his message pointing out an error in
the previous version of this announcement, and to
Prof.~D.~Yetter for his interest.  The author is also
indebted to Prof.  H.~D.~Doebner for his kind hospitality
in Clausthal.

\section{References}
{\frenchspacing
\item{1.} E. Witten, {\it Quantum Field Theory and the
Jones Polynomial}, Commun. Math. Phys. {\bf 121} (1989),
351--399.

\item{2.} N. Reshetikhin and V. G. Turaev, {\it Invariants
of 3-manifolds via link polynomials and quantum groups},
Invent. math. {\bf 103} (1991), 547--597.

\item{3.} R. Kirby and P. Melvin, {\it The 3-manifold
invariants of Witten and Reshetikhin-Turaev for sl(2,C)},
Invent. math. {\bf 105} (1991), 473--545.

\item{4.} W. B. R. Lickorish, {\it The skein method for
three-manifold invariants}, 1992 (unpublished).

\item{5.} M. H. Freedman and F. Luo, {\it Selected
Applications of Geometry to Low-Dimensional Topology},
University Lecture Series, American Mathematical Society,
Providence, 1989, Chapt. 4.

\item{6.} E. C\'esar de S\'a, {\it A link calculus for
4-manifolds}, Topology of low-dimensional manifolds, Proc.
Second Sussex Conf., Lecture Notes in Math., vol. {\bf
722}, Springer, Berlin, 1979, 16--30.

\item{7.} C. Blanchet, {\it Invariants on three-manifolds
with spin structure}, Comment. Math. Helvetici {\bf 67}
(1992), 406--427.

\item{8.} H. Ooguri, {\it Topological Lattice Models in
Four Dimensions}, Kyoto preprint RIMS-878 and e-preprint
hep-th/9205090.

\item{} L. Crane and D. Yetter, {\it A categorical
construction of 4D topological quantum field theories},
Kansas preprint and e-preprint hep-th/9301062.

\bigskip
}

{\cs
Department of Theoretical Physics, University of \L\'od\'z,
Pomorska 149/153, PL--90-236 \L\'od\'z, Poland
}

{\it Current address\/}:
Arnold Sommerfeld Institute for Mathematical Physics,
Technical University of Clausthal, Leibnizstra\ss e 10,
D-W--3392 Clausthal-Zellerfeld, Federal Republic of Germany

{\it E-mail address\/}:
ptbb@ibm.rz.tu-clausthal.de

\bye